\documentclass[%
 aip,
 amsmath,amssymb,
 reprint,longbibliography,
 citeautoscript, 
]{revtex4-1}
\usepackage[utf8]{inputenc}
\usepackage{graphicx}
\usepackage{bm}
\usepackage{bbold}
\usepackage{cases}
\usepackage{upgreek}
\usepackage[usenames, dvipsnames]{color}
\usepackage[colorlinks=true, citecolor=BrickRed, linkcolor=MidnightBlue, urlcolor=MidnightBlue ]{hyperref}
\usepackage{braket}
\usepackage[normalem]{ulem}
\usepackage{xfrac}

\usepackage{textgreek}
\usepackage{gensymb}
\usepackage{dcolumn}

\newcommand \mc[1] { \mathcal{#1} }
\newcommand \dd[1]  { \!\textrm d{#1} \,}   % infintesimal
\newcommand \rmm[1]  { \textrm{#1} }

\newcommand \e[1] { \rmm{e}^{#1} }

\newcommand \trb[1] { \rmm{tr}_\rmm{B}\left\{  #1  \right\} }
\newcommand \tr[1] { \rmm{tr}\left\{  #1  \right\} }
\newcommand \trs[1] { \rmm{tr}_\rmm{sys}\left\{  #1  \right\} }

\newcommand \ris { \!\!\! }
\newcommand \imi {\rmm{i}}

\makeatletter
\def\@email#1#2{%
 \endgroup
 \patchcmd{\titleblock@produce}
  {\frontmatter@RRAPformat}
  {\frontmatter@RRAPformat{\produce@RRAP{*#1\href{mailto:#2}{#2}}}\frontmatter@RRAPformat}
  {}{}
}%
\makeatother
\begin{document}

\preprint{AIP/123-QED}

\title{Generalized quantum master equations can improve the accuracy of semiclassical predictions of multitime correlation functions}

\author{Thomas Sayer}
\affiliation{Department of Chemistry, University of Colorado Boulder, Boulder, CO 80309, USA\looseness=-1}

\author{Andr\'{e}s Montoya-Castillo}
\homepage{Andres.MontoyaCastillo@colorado.edu}
\affiliation{Department of Chemistry, University of Colorado Boulder, Boulder, CO 80309, USA\looseness=-1} 

%--------------------------------------------------------------------------------------------

\date{\today}

\begin{abstract}

Multitime quantum correlation functions are central objects in physical science, offering a direct link between experimental observables and the dynamics of an underlying model. While experiments such as 2D spectroscopy and quantum control can now measure such quantities, the accurate simulation of such responses remains computationally expensive and sometimes impossible, depending on the system's complexity. A natural tool to employ is the generalized quantum master equation (GQME), which can offer computational savings by extending reference dynamics at a comparatively trivial cost. However, dynamical methods that can tackle chemical systems with atomistic resolution, such as those in the semiclassical hierarchy, often suffer from poor accuracy, limiting the credence one might lend to their results. By combining work on the accuracy-boosting formulation of semiclassical memory kernels with recent work on the multitime GQME, here we show for the first time that one can exploit a multitime semiclassical GQME to dramatically improve both the accuracy of coarse mean-field Ehrenfest dynamics and obtain orders of magnitude efficiency gains.

\end{abstract}

\maketitle

%\section{Introduction}

Multitime correlation functions, $\mc{C}(t_1, t_2, \dots, t_j)$, are fundamental dynamical objects that characterize chemical and physical systems. When a system is subject to external stimuli, its response can be formulated as a multitime correlation that relates directly to experimental observables, as in multidimensional spectroscopy\cite{Mukamel1999a} and quantum control protocols\cite{Viola1999, Brif2010, Shapiro2012}. Through this formalism the energetic structure and dynamics of all manner of physical systems become available: from nanocrystals to extended solids, from gas phase molecules to solvated chromophores. Beyond materials science, multitime correlation functions also reveal the extent of quantum information scrambling in model systems,\cite{Sadhasivam2023} chemical reactions,\cite{Zhang2022a, Zhang2024} black holes,\cite{Shenker2014} and many body-localized systems\cite{He2017} via out-of-time-ordered correlators. Hence, developing accurate and efficient theoretical and simulation frameworks to predict these objects is of paramount importance in chemistry, materials science, condensed matter physics, and astrophysics. 

Mathematically, while numerically exact solvers can predict the dynamics of few-state systems in contact with Gaussian heat baths,\cite{Tanimura1988, Meyer1990, Makri1995a, Thoss2001a, Wang2003a, Ishizaki2005, Suess2014, Tamascelli2019} going beyond effective harmonic environments and accounting for non-Gaussian, strong coupling to anharmonic motions remains a fundamental challenge. Such non-Gaussian interactions abound, emerging in: charge transfer within structured environments,\cite{Small2003a, Waskasi2018a, Matyushov2018a, Limmer2013a} optical spectroscopy of liquids,\cite{Bredenbeck2005a, Hamm2006a} interacting spin systems,\cite{Packwood2011, Degen2017a} shot noise in measurements of current\cite{Blanter2000, Huard2007} and radiation pressure\cite{Purdy2013}. Hence, a dynamical method should ideally capture faithful representations of fermion- and spin-nuclear couplings at the \textit{ab initio} level. Path integrals and semiclassical methods can, in principle, achieve this. However, in practice, these methods must balance accuracy and efficiency. Indeed, some techniques are accurate but prohibitively expensive, while others are largely inaccurate but (somewhat) affordable. It is thus crucial to develop reliable tools to reduce the computational cost and improve the accuracy of such schemes. 

Over the past decade, the generalized quantum master equation (GQME) framework\cite{Nakajima1958a, Zwanzig1960, Mori1965b} has proven transformative in this regard for 1-time correlation functions: GQMEs can both drastically reduce the cost of path integral, semiclassical, and classical simulations---sometimes by multiple orders of magnitude---and also boost the accuracy of even starkly incorrect semiclassical results to within graphical accuracy of benchmark calculations. Such improvements have been shown for the \textit{1-time} nonequilibrium and equilibrium dynamics of Gaussian\cite{Shi2003a, Shi2004b, Zhang2006, Kelly2015a, Kelly2013a, Montoya-Castillo2016, Kelly2016, Mulvihill2019b, Mulvihill2019c, Ng2021}, fermionic,\cite{Cohen2013, Cohen2013b, Kidon2015, Kidon2018a} and atomisitic\cite{Pfalzgraff2015a} models of charge and energy transfer in the condensed phase, and even for many-level photosynthetic complexes.\cite{Pfalzgraff2019a, Mulvihill2021c} The GQME can even be constructed using mean-field theory (MFT) simulations,\cite{Kelly2015a, Montoya-Castillo2016, Montoya-Castillo2017, Pfalzgraff2019a, Mulvihill2019b, Mulvihill2019c} arguably the least accurate semiclassical method. 

The GQME exactly rewrites a high-dimensional dynamics problem as a low-dimensional equation of motion of $\bm{\mc{C}}(t)$, which encodes the correlation of the variables of interest. At the heart of the GQME is the choice of the projection operator,\cite{Fick1990, Grabert2006} which accounts for the initial preparation of the chemical system (i.e., equilibrium vs.~nonequilibrium) and selects a few variables to track dynamically. The projection operator also defines the memory kernel, $\bm{\mc{K}}(t)$, which encapsulates the effect of all dynamical variables not being explicitly tracked and places an upper limit on how much simulation time is required to predict the dynamics of $\bm{\mc{C}}(t)$. The efficiency gains arise from adequately choosing a projection operator that minimizes the memory kernel lifetime. What leads to the accuracy gains is less clear. While early work posited that the accuracy improvements arise from the short-time accuracy combined with short-lived memory kernels, later work\cite{Montoya-Castillo2016} showed one could obtain efficiency gains even in memory kernels that are as long-lived as $\bm{\mc{C}}(t)$. In fact, Ref.~\onlinecite{Montoya-Castillo2016} proposed that the improved accuracy in 1-time nonequilibrium averages and correlation functions arises from the \textit{exact} semiclassical sampling of bath operators at $t=0$, which corrects the memory kernel through its self-consistent extraction from bath-augmented auxiliary kernels. 

We have recently shown that one can formulate GQMEs that drastically reduce the cost of multitime simulations,\cite{Sayer2024} but is it also possible to use these to improve \textit{semiclassical} multitime simulations? After all, recent advances in semiclassical theory have enabled the prediction of multitime correlation functions,\cite{Jansen2006, Liang2012a, Tempelaar2013a, VanDerVegte2013a, Loring2017a, Provazza2018b, Mannouch2022, Loring2022, Begusic2022} and the GQME could offer the ability to leverage this technology at a lower cost and greater accuracy. Yet, this is not a foregone conclusion. After all, our presumed source of improved accuracy, i.e., the \textit{exact} semiclassical sampling of bath correlations, is only strictly true at $t=0$. This raises important and revealing questions: 
\begin{enumerate}
    \item Does a bath-sampling GQME based on approximate semiclassical dynamics allow us to improve the accuracy across external temporal interactions? 
    \item How does additional bath sampling at later times impact the accuracy of the GQME predictions?
\end{enumerate}

To address these questions, we first describe how one must formulate 1-time semiclassical GQMEs to improve efficiency \textit{and} accuracy. We then introduce our extension of this method to 2-time simulations. We employ a spin-boson model with weak system-bath coupling and a fast decorrelating bath to compare our MFT and GQME results with an exact HEOM calculation. We refer the reader to Appendix~\ref{app:comp} for computational details.

%\section{Theory}
%\vspace{-4pt}

%\subsection{What makes the semiclassical 1-time GQME work?}
%\vspace{-10pt}

\textit{What makes the semiclassical 1-time GQME work?} In the Mori-Nakajima-Zwanzig GQME formalism\cite{Nakajima1958a, Zwanzig1960, Mori1965b}, one can obtain a low-dimensional ($N \times N$) equation of motion for the correlation function 
\begin{equation}
    \bm{\mc{C}}(t) = (\bm{A}\vert \e{-\imi \mc{L}t} \vert\bm{A}),
\end{equation}
$\bm{A}$ is a vector of $N$ operators whose dynamics we aim to track and the inner product can be defined as needed for the physical problem of interest. Using the projection operator, $\mathcal{P} = \vert\bm{A})(\bm{A}\vert$, yields the following time-nonlocal equation of motion for $\bm{\mc{C}}$,
\begin{equation}\label{eq:GQME}
    \dot{\bm{\mc{C}}}(t) = \dot{\bm{\mc{C}}}(0)\bm{\mc{C}}(t) - \int_0^t \dd{s} \bm{\mc{K}}(s)\bm{\mc{C}}(t-s).
\end{equation}
Here, we use the Argyres-Kelley projector, such that $\bm{A} = \{\vert 0\rangle\langle 0 \vert, \vert 0\rangle\langle 1 \vert, \vert 1\rangle\langle 0 \vert, \vert 1\rangle\langle 1 \vert\}$, and the inner product is the full trace over spectroscopically separable initial conditions, $\bm{\mc{C}}_{ij}(t) = \tr{A_j \e{-\imi\mc{L}t}A_i^\dagger\rho_B}$, where $\rho_B$ is the canonical density of the nuclear bath (see Appendix~\ref{app:comp}).

The GQME shifts the attention from the correlation function $\bm{\mc{C}}(t)$ to the memory kernel $\bm{\mc{K}}(t)$. That is, as long as one knows $\bm{\mc{K}}(t)$, it is easy to integrate Eq.~\ref{eq:GQME} to obtain the dynamics of $\bm{\mc{C}}(t)$. Generally, this memory kernel has a shorter lifetime, $\tau_\mc{K}$, than the correlation function's equilibration time, $t_\rmm{eq}$, setting an upper limit on the amount of information required to characterize the evolution of $\bm{\mc{C}}(t)$. That is $\tau_\mc{K} < t_\rmm{eq}$ where $\bm{\mc{K}}(t>\tau_\mc{K})=\bm{0}$.

The 1-time memory kernel takes the form, 
\begin{equation}
    \bm{\mc{K}}(t) = (\bm{A}\vert \mc{LQ} \e{-\imi \mc{QL}t} \mc{QL} \vert\bm{A}) ,
\end{equation}
where $\mc{Q} = \mathbb{1}-\mc{P}$ is the operator that projects objects in the full space onto the complementary space. However, the presence of the \textit{projected} propagator, $\e{-\imi \mc{QL}t}$, makes solving for the memory kernel challenging. While much work in quantum dynamics has focused on obtaining perturbative expansions of $\bm{\mc{K}}(t)$, we adopt its exact, self-consistent expansion\cite{Shi2003a, Montoya-Castillo2016, Kelly2016}
\begin{equation}\label{eq:Kauxiliary}
    \bm{\mc{K}}(t) = \bm{\mc{K}}^{(1)}(t) + \imi\int_0^t \dd{s} \bm{\mc{K}}(t-s)\bm{\mc{K}}^{(3b)}(s),
\end{equation}
into auxiliary kernels
\begin{equation}
    \bm{\mc{K}}^{(1)}(t) = (\bm{A}\vert \mc{LQ} \e{-\imi \mc{L}t} \mc{QL} \vert\bm{A})
\end{equation}
and
\begin{equation}
    \bm{\mc{K}}^{(3b)}(t) = (\bm{A}\vert \e{-\imi \mc{L}t}  \mc{QL} \vert\bm{A}),
\end{equation}
that require evolution with the \textit{full} propagator, $\e{-\imi \mc{L}t}$, instead. This expansion enables researchers to choose a convenient dynamical method to simulate the auxiliary kernels, compute $\bm{\mc{K}}(t)$, and ultimately predict $\bm{\mc{C}}(t)$. The superscript notation $\bm{\mc{K}}^{(3b)}$ is chosen to be consistent with previous work in the Heisenberg picture.\cite{Montoya-Castillo2016}

When one is interested in employing the GQME framework exclusively to improve computational efficiency, one can re-express the auxiliary kernels in terms of $\bm{\mc{C}}(t)$ and its time derivatives.\cite{Montoya-Castillo2016, Kelly2016} To obtain such expressions, one substitutes the definition of $\mc{Q}$ in the auxiliary kernels and replaces all Liouvillians with the numerical time derivatives. For example,
\begin{equation}
\begin{split}
    \bm{\mc{K}}^{(3b)}(t) &= (\bm{A}\vert  \e{-\imi \mc{L}t} \mc{L} \vert\bm{A}) - (\bm{A}\vert  \e{-\imi \mc{L}t} \vert\bm{A})(\bm{A}\vert \mc{L} \vert\bm{A}) \\
    &= \dot{\bm{\mc{C}}}(t) -  \bm{\mc{C}}(t)\dot{\bm{\mc{C}}}(0).
\end{split}
\end{equation}
Alternatively, one can use the tensor transfer method,\cite{Cerrillo2014a} which is formally equivalent. Thus, all relevant quantities can be extracted from reference dynamics for $\bm{\mc{C}}(t)$. 

For the GQME framework to improve the accuracy of \textit{1-time} semiclassical and quantum-classical dynamics, one must avoid fully decomposing the auxiliary kernels into $\bm{\mc{C}}(t)$ and its time derivatives. Exceptions arise when using methods whose approximate propagator, $\e{-\imi \mc{L}t}_{\rm app}$, commutes with the Liouvillian, $\mathcal{L}$, or that preserve the canonical distribution when computing equilibrium correlation functions. In such cases, the GQME framework cannot lead to any improvement in accuracy,\cite{Kelly2016} but can still be used to improve computational efficiency. That is, the quality of the dynamics obtained is the same as that of the original semiclassical method used to parameterize it. However, since these exceptions do not arise for most methods in the semiclassical hierarchy\cite{Heller1975, Herman1994, Stock1997a, Sun1997b, Shao1999, Thoss2004b, Bonella2005a, Miller2009a, Huo2011a, Richardson2013a, Ananth2022} or that arise from the quantum-classical Liouville equation,\cite{McLachlan1964a, Tully1990a, Stock1995a, Hammes-Schiffer1996, Kapral1999a, Shi2004c, Kim2008a, Hsieh2012a, Kapral2015a, Subotnik2016, Runeson2023} one may expect a suitable formulation of the semiclassical GQME to lead to improved dynamics.

\begin{figure}[!t]
\vspace{-2pt}
\begin{center} 
    \resizebox{.45\textwidth}{!}{\includegraphics{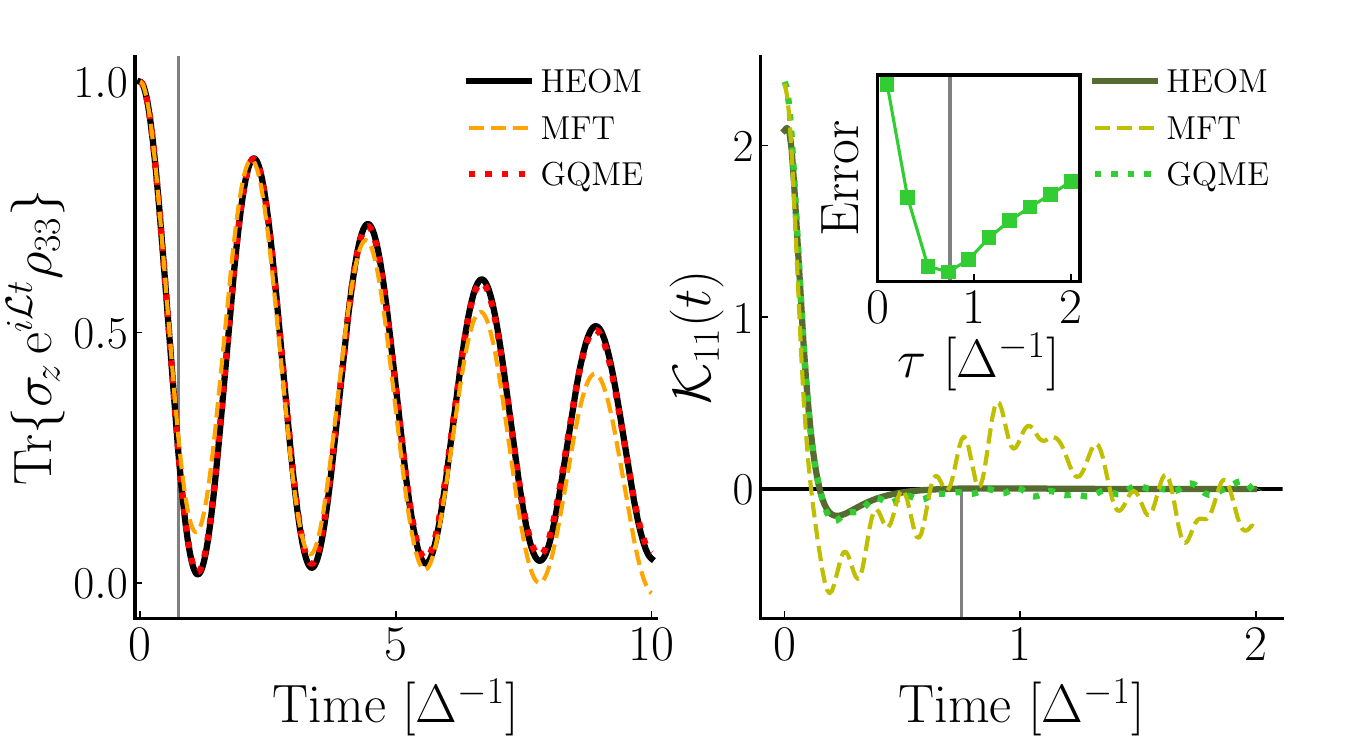}}
\vspace{-6pt}
\caption{\label{fig:onetime} HEOM, MFT, and GQME results for a spin-boson model with $\Delta = \epsilon$, $\omega_c = 7.5\Delta$, $\lambda = 0.075\Delta$, $\beta = 1/\Delta$ and an Ohmic spectral density. The grey line marks the cutoff, 0.7~$\Delta^{-1}$, used to generate the GQME dynamics. \textbf{Left}: Difference between site populations after initializing on the lower-energy site. The GQME (red dotted) recovers the HEOM reference (black), in contrast to the direct MFT result (dashed yellow). \textbf{Right}: Representative element of the memory kernel for all methods over the first 2~$\Delta^{-1}$. \textbf{Inset}: The GQME error given truncation of the kernel (dotted green) at time $\tau$. Error increases after a minimum due to the poorer convergence of the MFT trajectories at later times.}
\end{center}
\vspace{-20pt}
\end{figure}

In anticipation of the multitime derivation, and to contextualize the extent of GQME-based improvement that one may expect in the multitime response, we first illustrate how the semiclassical GQME can improve the accuracy of the MFT electronic dynamics in a parameter regime corresponding to a biased, underdamped spin connected to a fast decorrelating bath. \mbox{Figure~\ref{fig:onetime}} compares the performance of the MFT against benchmark HEOM calculations and illustrates the improvement one obtains from the GQME sampling bath correlations during the MFT simulation. The GQME brings the approximate MFT dynamics to within graphical accuracy of the exact HEOM. We have chosen this parameter regime as it places particular strain on MFT: like many other semiclassical methods, MFT violates detailed balance, a flaw that becomes most evident in biased systems; and the classical approximation for the bath deteriorates the most for fast, high-frequency baths. While a lower temperature would have rendered the classical treatment of the bath in MFT more questionable and heightened the disagreement between MFT and HEOM, anticipating a nearly prohibitive cost of converging the multitime HEOM benchmark we have adopted an intermediate temperature that allows faster convergence with the number of Matsubara frequencies. These data demonstrate why the argument claiming that accuracy improvement arises from the short-time accuracy of MFT (leading to memory kernels that are accurate over their short lifetimes) is incomplete. Indeed, the GQME kernel is more accurate than the MFT equivalent even before the first minimum in the dynamics at very early times. 

At the technical level, the GQME displayed in Figure~\ref{fig:onetime} circumvents decomposing the auxiliary kernels into $\bm{\mc{C}}(t)$ and its time derivatives and instead directly simulates the \textit{new} correlation functions one obtains by rotating the initial and final conditions in the auxiliary kernels. That is, $\mc{QL}|\mathbf{A}) = \mc{L}_{SB}|\mathbf{A})$ and $(\mathbf{A}|\mc{LQ} = (\mathbf{A}|\mc{L}_{SB}$, where $\mc{L}_{SB} = [H_{\rm SB}, \dots]$ is the system-bath Liouvillian (see Appendix~\ref{app:kernel}). For example, noting that,
\begin{equation}\label{eq:Hsb}
    \hat{H}_{\rm SB} = \hat{\sigma}_z \sum_{n} c_n \hat{q}_n \equiv \hat{\sigma}_z \otimes V_B,
    \vspace{-4pt}
\end{equation}
one finds that for the row $n$ of $\bm{\mc{K}}_{nm}^{(3b)}(t)$, one needs to rotate the initial condition as follows:
\begin{equation}
\begin{split}
    \mc{L}_{SB}(\rho_B\otimes \hat{A}_n^\dagger) &= \rho_B\mc{L}_{SB}(\hat{A}_n^\dagger) + \hat{A}_n^\dagger \mc{L}_{SB}(\rho_B) \\
    &= \rho_BV_B [\sigma_z, \hat{A}_n^\dagger] + \hat{A}_n^\dagger \sigma_z [V_B, \rho_B].
\end{split}
\end{equation}

\unskip\parfillskip 0pt Thus, calculating this quantity without using the time derivatives of $\bm{\mc{C}}(t)$ requires sampling the bath coupling $V_B$ (here at time zero) in addition to the original system observables, $\ket{j}\bra{k}$. Furthermore, to sample the resulting correlation functions semiclassically, one must { \par}

\pagebreak \noindent Wigner transform the additional bath operators.\cite{Kapral2015a} Since the functional forms require capturing the Wigner transform of products of bath operators, one must employ the Moyal bracket,\cite{Imre1967a, Hillery1984}
\vspace{-4pt}
\begin{equation}\label{eq:K3bWignerEg}
    \begin{split}
        (\rho_B V_B)_w &= \rho_{B,w} \e{\hbar\Lambda/2\imi} V_{B,w} \\
        &= \rho_{B,w} (1+\hbar\Lambda/2\imi) V_{B,w} \\
        &= \rho_{B,w}V_{B,w} + \imi\rho_{B,w}\zeta_{B,w},
    \end{split}
\end{equation}
where $\Lambda$ is the classical Poisson bracket and the linearization is exact when $V_B$ is a linear function of the bath coordinates. This shows that the Wigner transformation that one uses to facilitate a semiclassical treatment generates an extra term, $\zeta_{B,w}$, containing the derivative of the coupling (see Appendix~\ref{app:comp}). For a term like $\mc{K}_{nm}^{(1)}(t)$, performing all rotations arising from $\mathcal{L}_{SB}$ requires sampling bath operators also at time~$t$. It turns out that only the time-zero sampling is significant to obtaining the true kernel \cite{Montoya-Castillo2016}, but here we work with the fully rotated expressions and test this observation explicitly, later.

%\subsection{Semiclassical Multitime GQME}
\textit{Semiclassical multitime GQME:} We now consider the case where the system experiences an impulsive external field at time $t_1$, which we choose to be the $\sigma_x$ Pauli matrix. This corresponds to a photon-induced spin-flip of a qubit in quantum information protocols.\cite{Viola1999} As the interaction produces a state that cannot be separated into a system-bath product state, the evolution after $t_1$ does not follow the 1-time GQME equation that held before the impulse. Instead, we must invoke a multitime GQME. This formalism expresses the correction term using a new memory kernel-like object that takes $t_1$ as an argument. We refer to it as the 2-time kernel $\bm{\mc{K}}^{\sigma_x}(t_2, t_1)$. 
% This is a two-time problem, but the derivation we provide is trivially extended to higher orders.

In the multitime GQME formalism,\cite{Ivanov2015, Sayer2024} one can decompose the 2-time correlation function as follows
\begin{widetext}
    \begin{equation}\label{eq:2timeprop}
    \bm{\mc{C}}^{\sigma_x}(t_2,t_1) = \bm{\mc{C}}(t_2)\bm{\Sigma_x}\bm{\mc{C}}(t_1) + \int_0^{t_2}\!\int_0^{t_1}\dd{\tau_2}\dd{\tau_1}\bm{\mc{C}}(t_2-\tau_2)\bm{\mc{K}}^{\sigma_x}(t_2, t_1)\bm{\mc{C}}(t_1-\tau_1),  
    \end{equation}
    \vspace{-10pt}
\end{widetext}  
where $\bm{\Sigma_x}$ is the matrix $(\bm{A}\vert \hat{\sigma}_x \vert \bm{A})$. The first term can be constructed using only 1-time data and represents the solution when the entanglement induced by the measurement at $t_1$ is ignored. The second term isolates this 2-time contribution and requires a full 2-time simulation to obtain. For the parameter regime in Fig.~\ref{fig:onetime}, we expect the 2-time entanglement term to be small\cite{Sayer2024} but otherwise representative of the protocol we are delineating. The 2-time kernel subject to a  $\hat{\sigma}_x$ interaction at $t_1$ is 
\vspace{-4pt}
\begin{equation}\label{eq:2timekernel}
    \bm{\mc{K}}^{\sigma_x}(t_2,t_1) = (\bm{A}|\mc{L} \e{-\imi\mc{Q}\mc{L}t_2} \mc{Q} \hat{\sigma}_x \e{-\imi\mc{Q}\mc{L}t_1} \mc{Q}\mc{L} |\bm{A}). \vspace{-2pt}
\end{equation}
We have previously shown how to self-consistently expand this multitime kernel and decompose it into auxiliary kernels that employ only derivatives of the parent 2- and 1-time correlation functions.\cite{Sayer2024} However, such decompositions cannot offer accuracy improvements when combined with semiclassics. Instead, in Appendix~\ref{app:kernel}, we show how one could obtain expressions that can offer accuracy improvements by explicitly considering the $\mc{L}_{SB}$-induced rotations in the auxiliary kernels. The first difference with our previous work is that instead of derivatives of the dynamics appearing under the convolution integrals (as in Eq.~\ref{eq:Kauxiliary}), we require the semiclassical, bath-sampled 1-time $\bm{\mc{K}}^{(3f)}(t)$ and $\bm{\mc{K}}^{(3b)}(t)$. The second difference is that we must Wigner transform the 2-time equivalent of $\bm{\mc{K}}^{(1)}$,
\begin{widetext}
    \vspace{-16pt}
    \begin{align} \label{eq:KsigmaWigner}
        (\bm{A}|L_{SB}\e{-\imi\mc{L}t_2}\hat{\sigma}_x\e{-\imi\mc{L}t_1}\mc{L}_{SB}|\bm{A})_{nm} &= \iint \dd{\mathbf{p}}\,\dd{\mathbf{q}} \rho_{B,w}V_{B,w}V_{B,w}(t_1+t_2) X_{ni}\trs{\hat{A}_i^\dagger \hat{\sigma}_x(t_1) \hat{A}_j(t_1+t_2)}X_{jm} \nonumber\\
        &~~~~~- \iint \dd{\mathbf{p}}\,\dd{\mathbf{q}} \rho_{B,w}\zeta_{B,w}V_{B,w}(t_1+t_2) Y_{ni}\trs{\hat{A}_i^\dagger \hat{\sigma}_x(t_1) \hat{A}_j(t_1+t_2)}X_{jm},
    \end{align}
    \vspace{-10pt}
\end{widetext}
where $\bm{X}$ and $\imi\bm{Y}$ are diagonal rotation matrices representing the action of $\hat{\sigma}_z$ in an anticommutator and commutator, respectively.\cite{Montoya-Castillo2016} This shows that one must measure bath operators $V_{B,w}$ and $\zeta_{B,w}$ at $t=0$ and $V_{B,w}$ at $t=t_1+t_2$, in analogy to the 1-time case. There is no need to sample the bath variables during the intermediate evolution (at $t_1$) because $\mc{L}_{SB}$ only appears at the start and end of the inner product. The same protocol holds for higher-order multitime protocols.

%------------------------------------------------------------------

\begin{figure}[!b]
\vspace{-18pt}
\begin{center} 
    \resizebox{.45\textwidth}{!}{\includegraphics{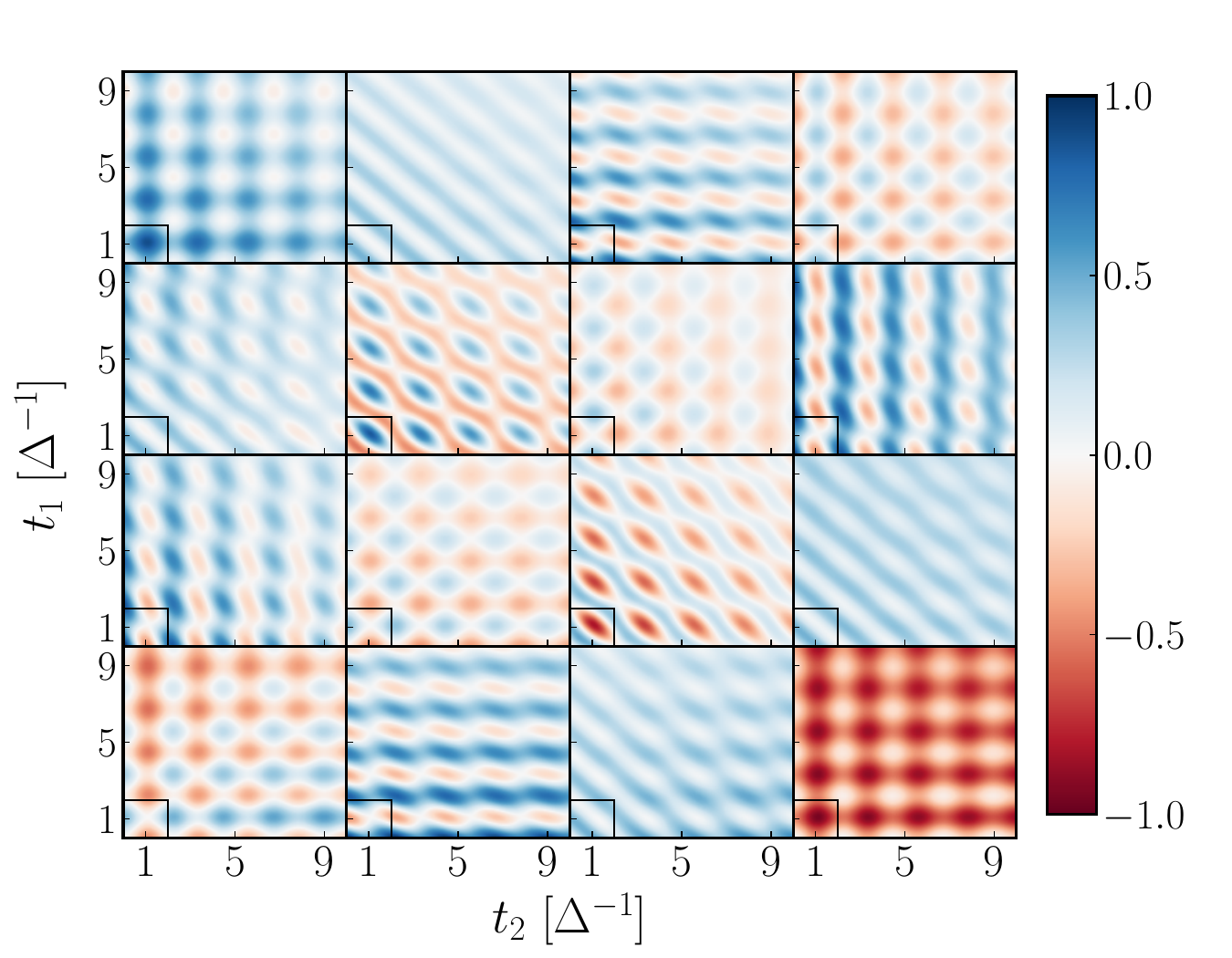}}
\vspace{-8pt}
\caption{\label{fig:heom_twotime} Two-time correlation function obtained using HEOM of our spin-boson model with $\epsilon = \Delta$, $\omega_c = 7.5\Delta$, $\lambda = 0.075\Delta$, $\beta = \Delta$ and an Ohmic spectral density. The total grid displays lightly damped oscillations also evident in Fig.~\ref{fig:onetime}. The black boxes in the lower left of each panel demarcate the time window we over which the 2-time memory kernels decay.}
\end{center}
\vspace{-24pt}
\end{figure} 

%results
We now turn to our simulation results. First, we compute the benchmark 2-time correlation function (Fig.~\ref{fig:heom_twotime}) using the numerically exact HEOM over a ${10~\Delta^{-1}\times10~\Delta^{-1}}$ grid of points with a $0.01$~$\Delta^{-1}$ resolution. For the GQME, we do not need to directly simulate the entire grid, which is one of the central benefits of the method. Instead, we demarcate the first 2~$\Delta^{-1}\times2$~$\Delta^{-1}$ of the grid with black boxes and first simulate over this region with MFT. We choose this value because our 1-time kernel lifetime is 0.7~$\Delta^{-1}$ (see Fig.~\ref{fig:onetime}), and our previous work \cite{Sayer2024} suggests the 2-time kernel goes to zero at least as fast as the 1-time kernel. A 2~$\Delta^{-1}$ range on both axes should be more than sufficient to fully parameterize the 2-time kernel, and thus generate the GQME dynamics for the full ${10~\Delta^{-1}\times10~\Delta^{-1}}$ grid in Fig.~\ref{fig:heom_twotime}. 

\begin{figure}[!b]
\vspace{-18pt}
\begin{center} 
    \hspace{-15pt}
    \resizebox{.25\textwidth}{!}{\includegraphics{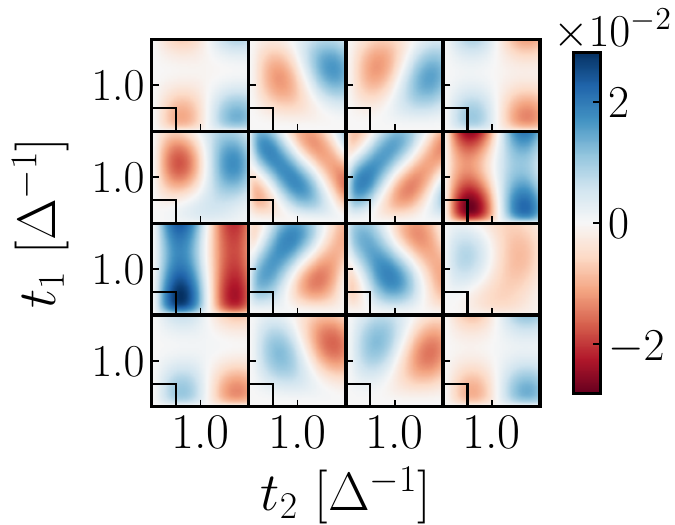}}
    \hspace{-5pt}
    \resizebox{.25\textwidth}{!}{\includegraphics[trim={20pt, 0, -20pt, 0},clip]{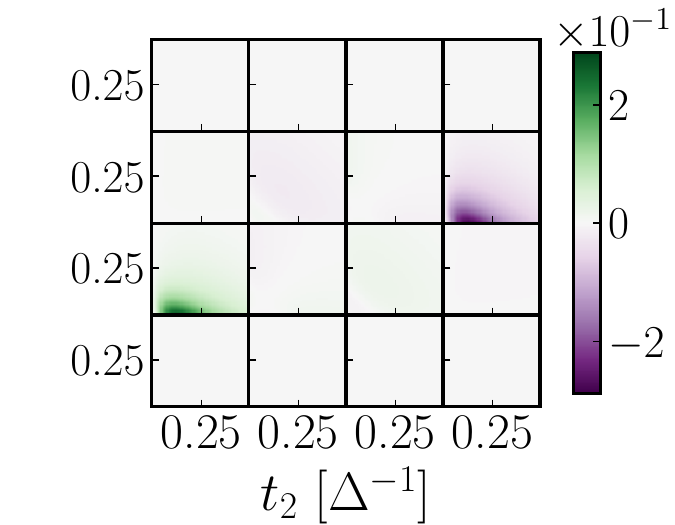}}
\vspace{-6pt}
\caption{\label{fig:heom_LHSandKA} From the HEOM data of Fig.~\ref{fig:gqme_twotime} we calculate \textbf{Left}: the 2-time correction to the 1-time approximation corresponding to the second term of Eq.~\ref{eq:2timeprop}, $\bm{\mc{C}}^{\sigma_x}(t_2,t_1) - \bm{\mc{C}}(t_2)\bm{\Sigma_x}\bm{\mc{C}}(t_1)$, and \textbf{Right}: the 2-time kernel, $\bm{\mc{K}}^{\sigma_x}(t_2,t_1)$, of Eq.~\ref{eq:2timekernel}. The 0.5~$\Delta^{-1}$ axis limits are shown as black boxes in the left panel.}
\end{center}
\vspace{-24pt}
\end{figure}

To quantify the improvement that the multitime semiclassical GQME can afford, we home in on the difference between the full propagator and the 2-time correlation function constructed using only 1-time quantities, i.e., the second term of Eq.~\ref{eq:2timeprop}. Figure~\ref{fig:heom_LHSandKA}--left shows the benchmark of this quantity obtained with HEOM. For these exact data, we employ a continuous-time algorithm (see Appendix~A of Ref.~\onlinecite{Sayer2024}) to extract the 2-time kernel, shown in Fig.~\ref{fig:heom_LHSandKA}--right from the 2-time term. In this regime, the 2-time kernel has only two significant elements that are confined to small triangles of $\sim 0.5$~$\Delta^{-1}$ base. This confirms our prediction that $2$~$\Delta^{-1}$ is an adequate simulation time for the MFT trajectories, to which we now turn.

\begin{figure}[!t]
\vspace{-4pt}
\begin{center} 
    \hspace{-15pt}
    \resizebox{.25\textwidth}{!}{\includegraphics{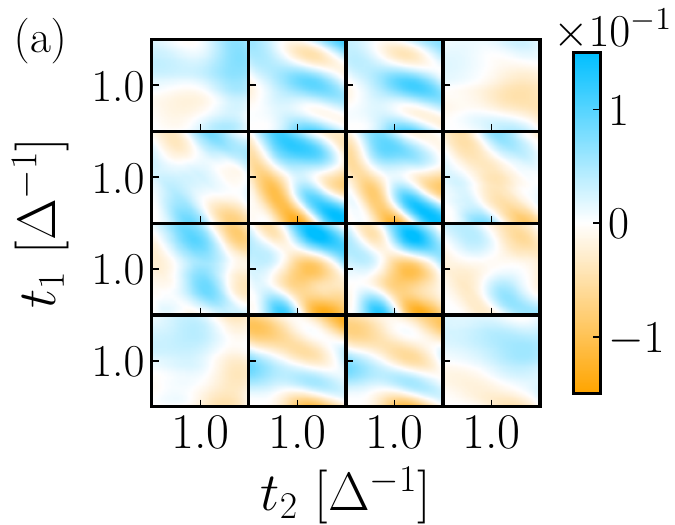}}
    \hspace{-5pt}
    \resizebox{.25\textwidth}{!}{\includegraphics[trim={20pt, 0, -20pt, 0},clip]{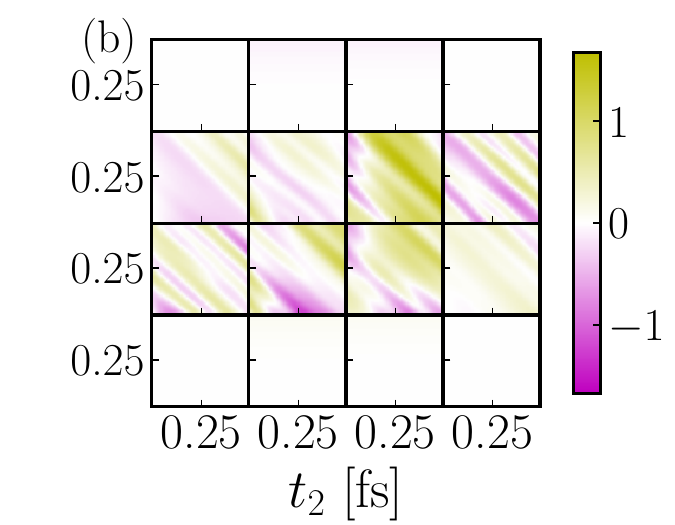}}
    \\
    \vspace{-14pt}
    \hspace{-15pt}
    \resizebox{.25\textwidth}{!}{\includegraphics{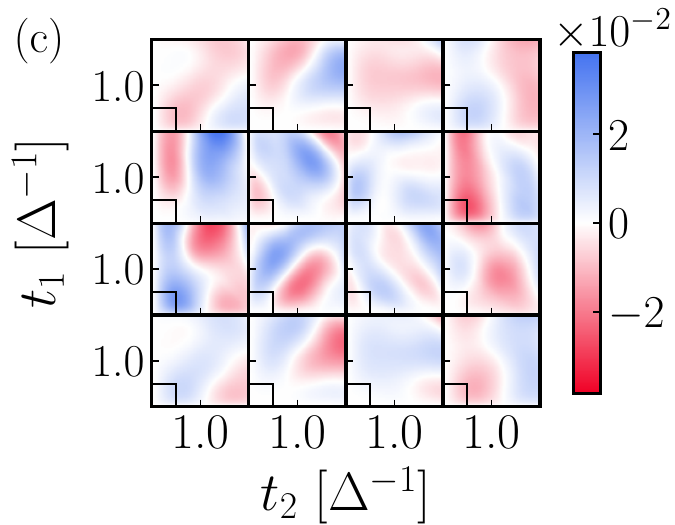}}
    \hspace{-5pt}
    \resizebox{.25\textwidth}{!}{\includegraphics[trim={20pt, 0, -20pt, 0},clip]{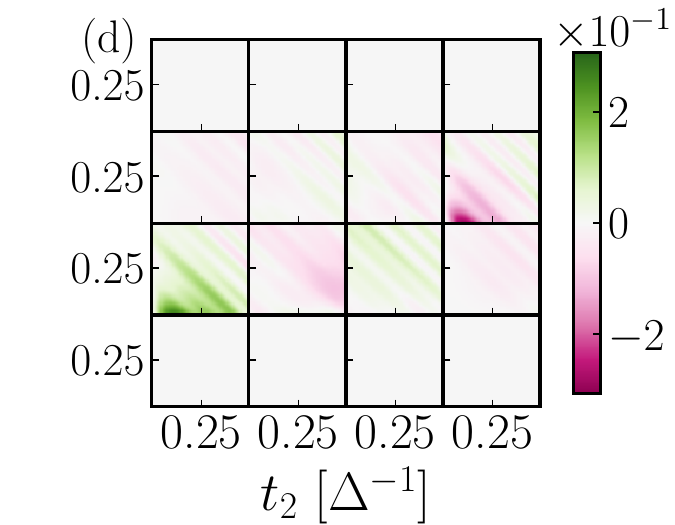}}
\vspace{-8pt}
\caption{\label{fig:ehre_LHSandKA} \textbf{Top}: MFT results for \textbf{(a)}: 2-time term $\bm{\mc{C}}^{\sigma_x}(t_2,t_1) - \bm{\mc{C}}(t_2)\bm{\Sigma_x}\bm{\mc{C}}(t_1)$ and \textbf{(b)}: 2-time kernel $\bm{\mc{K}}^{\sigma_x}(t_2,t_1)$ calculated using the same (exact) derivative relationships as in Fig.~\ref{fig:heom_LHSandKA}. \textbf{Bottom}: GQME results for \textbf{(c)}: 2-time term $\int_0^{t_2}\!\int_0^{t_1}\dd{\tau_2}\dd{\tau_1}\bm{\mc{C}}(t_2-\tau_2)\bm{\mc{K}}^{\sigma_x}(t_2, t_1)\bm{\mc{C}}(t_1-\tau_1)$ where the \textbf{(d)}: 2-time kernel is calculated with Eq.~\ref{eq:KsigmaWigner} and the improved 1-time quantities as described in Appendix~\ref{app:kernel}.}
\end{center}
\vspace{-18pt}
\end{figure}

Figure~\ref{fig:ehre_LHSandKA} presents both MFT and semiclassical GQME equivalents to the panels of Fig.~\ref{fig:heom_LHSandKA}. We employ the semiclassical GQME-improved version of the 1-time quantities presented in Fig.~\ref{fig:onetime}, which are effectively identical to the HEOM in this case. This enables us to isolate all errors in the 2-time term to the 2-time kernel. The top row of Fig.~\ref{fig:ehre_LHSandKA} shows that MFT is completely inaccurate, even at the shortest times. Specifically, the 2-time term of Fig.~\ref{fig:ehre_LHSandKA}a has no clear structure and is almost an order of magnitude too large. This means that, in this regime, discarding the 2-time term completely and just using 1-time improvements would be preferable to the result of the 2-time calculation. We construct the associated 2-time kernel in Fig.~\ref{fig:ehre_LHSandKA}b from derivatives of the 2-time term, and therefore obtain an object which is also of the wrong form and magnitude.

Interestingly, if one rotates only the first $\mc{QL}$ term with $\mc{L}_{SB}$ when obtaining Eq.~\ref{eq:KsigmaWigner}, i.e., performing bath sampling only at $t=0$, then the improvement is lost. Previous work \cite{Montoya-Castillo2016} found that when replacing the remaining $\mc{Q}$ with $\mathbb{1}-\mc{P}$, the difference between the two terms was slow to converge as a function of the number of trajectories. Hence, we ascribe this observation to the increased difficulty of converging the small difference of two large terms subject to statistical noise from finite sampling.

\begin{figure}[!b]
\vspace{-4pt}
\begin{center} 
    \hspace{-15pt}
    \resizebox{.25\textwidth}{!}{\includegraphics{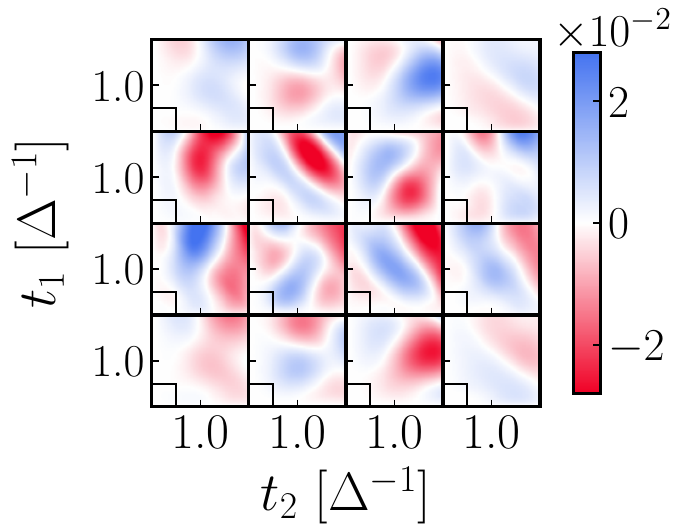}}
    \hspace{-5pt}
    \resizebox{.25\textwidth}{!}{\includegraphics[trim={20pt, 0, -20pt, 0},clip]{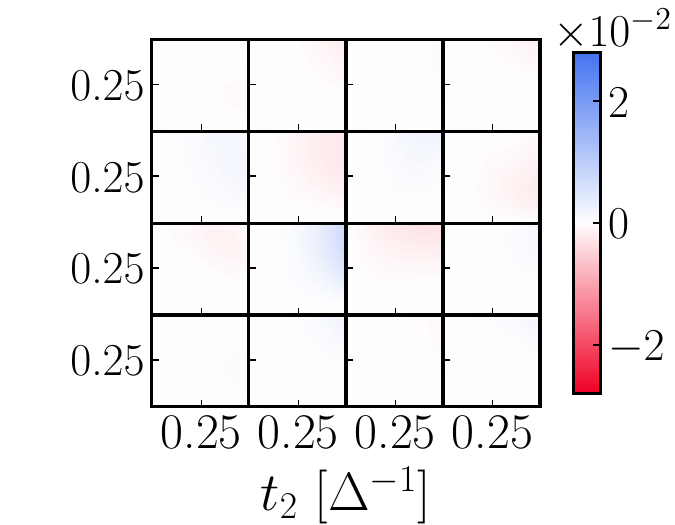}}
\vspace{-8pt}
\caption{\label{fig:GQME_diffs_zooms} Differences between the 2-time correction for the GQME, Fig.~\ref{fig:ehre_LHSandKA}c, and that of HEOM, Fig.~\ref{fig:heom_LHSandKA}--left, \textbf{Left}: over the full range of Fig.~\ref{fig:ehre_LHSandKA}c and \textbf{Right}: over the region needed to construct the kernel of Fig.~\ref{fig:ehre_LHSandKA}d, where the error is insignificant on the appropriate, $10^{-2}$ scale.}
\end{center}
\vspace{-12pt}
\end{figure}

Currently, the MFT multitime protocol is computationally expensive: our 2-time MFT simulations average over 10,000 trajectories, which has the same cost of performing 100,000,000 1-time trajectories. To obtain the 2-time analog of the 100,000 that satisfactorily converge the 1-time data in Fig.~\ref{fig:onetime} would take 4~orders of magnitude longer. While converging the data over the entire time domain is feasible for the spin-boson model as semiclassical calculations are trivially parallelizable, performing this many simulations with an atomistic bath would become difficult as force calculations constitute the greatest expense in molecular dynamics. For this reason, we demonstrate that it is possible to obtain improved accuracy and efficiency through our semiclassical GQME \textit{even without fully converging the long-time data}. Indeed, while at this level of convergence the 2-time kernel of Fig.~\ref{fig:ehre_LHSandKA}d is still noisy compared to the HEOM result in Fig.~\ref{fig:heom_LHSandKA}--right, the GQME 2-time term in Fig.~\ref{fig:ehre_LHSandKA}c obtains semiquantitative agreement in form and magnitude. Additionally, because it is a trajectory-based method, earlier time data is more converged, and we only need convergence in the region used in the memory kernel construction, i.e., the smaller boxes in Fig.~\ref{fig:ehre_LHSandKA}--left. To be quantitative about the degree to which the GQME kernel is underconverged, in Fig.~\ref{fig:GQME_diffs_zooms} we show the difference between GQME and HEOM 2-time terms on these two timescales.

\begin{figure}[!t]
\vspace{-8pt}
\begin{center} 
    \resizebox{.45\textwidth}{!}{\includegraphics{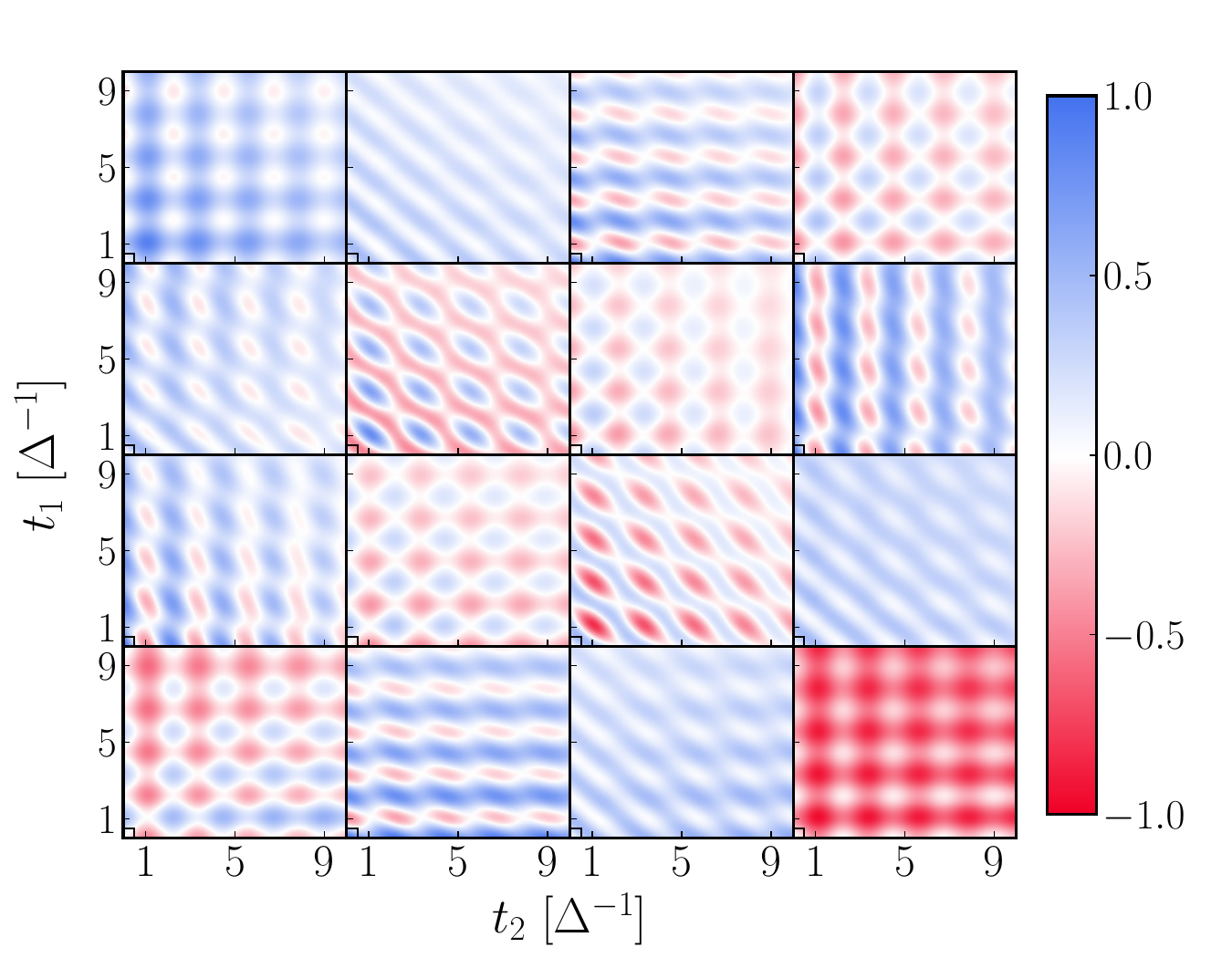}}
\vspace{-8pt}
\caption{\label{fig:gqme_twotime} 2-time correlation function from the semiclassical GQME, to be compared with the HEOM equivalent of Fig.~\ref{fig:heom_twotime}. In this figure, the (smaller) black boxes instead shows the region of time that needed to be simulated in order to compute the full grid.}
\end{center}
\vspace{-20pt}
\end{figure}

\pagebreak
The fact that there is little to no error in the 0.5~$\Delta^{-1}$ region of Fig.~\ref{fig:GQME_diffs_zooms}--right directly represents an improvement in accuracy: while the 2-time term is not maximal in this region, one can see from the $(1, 3)$ and $(2,0)$ elements of Fig.~\ref{fig:heom_LHSandKA}--left that it is non-zero and not trivial to obtain. The larger errors at later times are due to an accumulation of error from the underconverged tails of $\bm{\mc{K}}^{\sigma_x}(t_2, t_1)$, consistent with the observation semiclassical methods require increasingly larger numbers of trajectories to converge correlation functions evaluated at longer times. Since the size of the 2-time term itself is the same as the error, careful comparison shows this is mostly due to slightly shifted frequencies (relative positions of regions of positive and negative intensity) rather than their intensities. We can therefore have some confidence that---since the magnitudes are correct, the frequencies are largely accurate, and the GQME 2-time kernel of Fig.~\ref{fig:ehre_LHSandKA}c clearly goes to zero---it is reasonable to evolve the GQME to the full $10$~$\Delta^{-1}$$\times 10$~$\Delta^{-1}$ grid of Fig.~\ref{fig:heom_twotime}. Using just the data in the 0.5~$\Delta^{-1}$$\times 0.5$~$\Delta^{-1}$ region and the 1-time improvement (which is of insignificant cost in comparison) we obtain the GQME 2-time propagator and display it in Fig.~\ref{fig:gqme_twotime}. 

\begin{figure}[!t]
\vspace{-8pt}
\begin{center} 
    \resizebox{.45\textwidth}{!}{\includegraphics{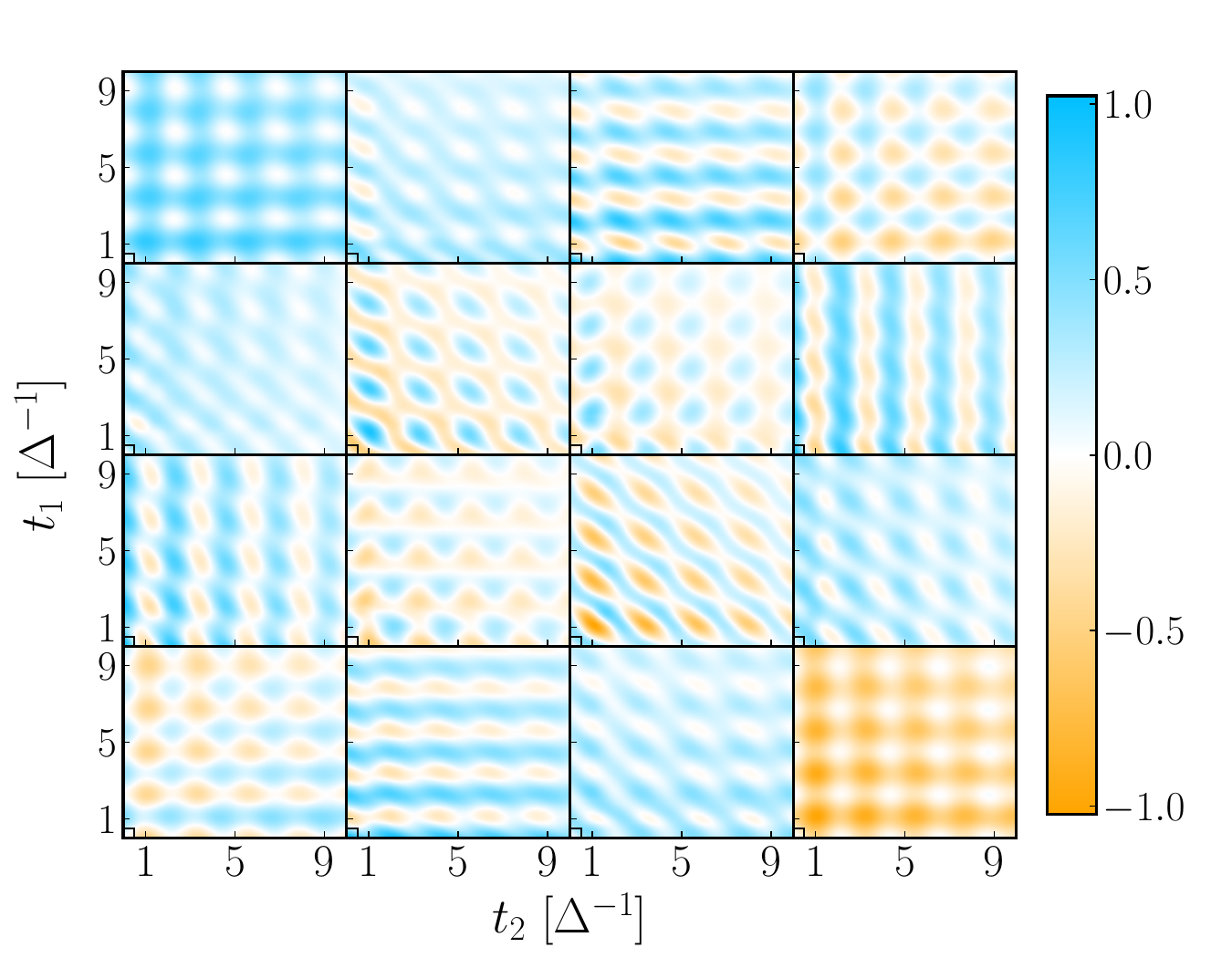}}
\vspace{-8pt}
\caption{\label{fig:MFT_twotime} GQME-extended 2-time correlation function using only the original MFT 2-time dynamics formulated to recover dynamics of MFT-level accuracy. The cutoff is 1~$\Delta^{-1}$. When compared with Figs.~\ref{fig:gqme_twotime}~and~\ref{fig:heom_twotime}, significant disagreement is clear in both frequencies and amplitudes, analogous to Fig.~\ref{fig:onetime}--left.}
\end{center}
\vspace{-15pt}
\end{figure}

The agreement between the HEOM and semiclassical GQME results for the multitime correlation function $\bm{\mc{C}}^{\sigma_x}(t_2, t_1)$ in Figs.~\ref{fig:gqme_twotime} and \ref{fig:heom_twotime} is excellent, even at later times. Compare, for example, the small red dots in the white `wells' of the $(0,0)$ element of both figures; these features represent, for example, the extent to which a quantum control sequence could be used to manipulate the coherence properties of this spin system. In contrast, Fig.~\ref{fig:MFT_twotime} shows the ``full MFT'' multitime correlation function. To generate this figure, we do not perform the 2-time MFT simulation over the $10$~$\Delta^{-1}$ $\times 10$~$\Delta^{-1}$ grid as converging it with the same resolution becomes prohibitively expensive. Instead, we employ both 1-time and 2-time GQME formulae from our previous multitime GQME work,\cite{Sayer2024} which exclusively improve the efficiency but not the accuracy of semiclassics. Features like these small dots are entirely missing in the 2-time MFT result and the qualitative structure of many elements is incorrect (e.g. the (2, 3) element has a pattern of wells where it should be diagonal stripes), showing how critical it is to sample bath information when employing a GQME with MFT reference dynamics. Also noteworthy is that our parameter regime corresponds to weak electronic-nuclear coupling (small reorganization energy), which leads our 2-time memory kernel to offer a comparatively small correction. However, many experimental systems display intermediate to strong electronic-nuclear coupling, such as light-harvesting complexes\cite{Mirkovic2017a, Jang2018b} and small polaron-forming materials such as polymers\cite{Ghosh2020a, Nematiaram2020a} and transition metal oxides\cite{Iordanova2005, Franchini2021}. In such cases, a 2-time kernel five times too large would lead to catastrophically incorrect 2-time propagators even with exact 1-time dynamics.

% \section{Conclusion}
% \vspace{-4pt}
In conclusion, we showed a representative example of when the MFT method cannot quantitatively or even qualitatively capture the 2-time correction to a quantum multitime correlation function reporting on light-induced interactions in the spin-boson model. We then demonstrated that our semiclassical multitime GQME dramatically improved the accuracy the 2-time kernel by employing classical sampling of bath interactions obtained from the same qualitatively and quantitatively incorrect MFT dynamics. Computationally, the semiclassical GQME gave a modest $10/0.7\simeq14$ times efficiency gain over the 1-time window, and impressive accuracy gains that brought the semiclassical MFT into agreement with the HEOM benchmark, consistent with previous work. For the 2-time problem, the $10/0.5=20$ times improvement was squared, meaning that to produce the figures in this paper, in which the dynamics have not yet even reached equilibrium, our multitime semiclassical GQME reduced the cost of the simulation by $400$~times. Just as impressively, extending the multitime simulation to equilibrium incurs only a trivial computational cost compared with the effort to do this with either HEOM or semiclassical theory alone. Our illustrative example also demonstrates a general principle that since the higher-order memory kernels have lifetimes comparable to (and even along one axis, shorter than) the 1-time kernel, the multitime GQME obtains higher savings as the number of time indices in the correlation function increases. This offers a promising approach for obtaining accurate and efficient semiclassical 2D spectra for complex systems.
%-----------------------------------------------------------------------
\vspace{-14pt}
\begin{acknowledgments}
\vspace{-8pt}
This work was partially supported by an Early Career Award in CPIMS program in the Chemical Sciences, Geosciences, and Biosciences Division of the Office of Basic Energy Sciences of the U.S.~Department of Energy under Award DE-SC0024154. We also acknowledge start-up funds from the University of Colorado Boulder.
\end{acknowledgments}

\vfill\pagebreak

% \section*{AUTHOR DECLARATIONS}
% \vspace{-12pt}
% \section*{Conflict of Interest}
% \vspace{-8pt}
% The authors have no conflicts to disclose.
\vspace{-12pt}
\section*{Author Contributions}
\vspace{-8pt}
Thomas Sayer: Formal analysis (lead); Investigation (lead);
Writing – original draft (lead); Writing – review \& editing (equal).
Andrés Montoya-Castillo: Conceptualization (lead); Supervision
(lead); Writing – review \& editing
(equal).
\vspace{-12pt}
\section*{DATA AVAILABILITY}
\vspace{-8pt}
{\raggedright The data that support the findings of this study are available from the corresponding author upon reasonable request. }

\vspace{-12pt}
\appendix

\section{Computational Details}\label{app:comp}
\vspace{-10pt}
We performed all HEOM calculations using the open-source pyrho code \cite{Berkelbach2020}, and 2-time Ehrenfest simulations with an in-house implementation of the pure-state protocol introduced in Ref.~\onlinecite{Atsango2023}.

In this paper, we choose a biased regime of $\epsilon = \Delta$ with a fast bath $\omega_c = 7.5\Delta$, and weak coupling $\lambda = 0.075\Delta$, at an intermediate temperature of $\beta = 1/\Delta$. We use an Ohmic spectral density which converges the MFT faster than the other common choice (Debye). We choose this regime because the memory kernel is fairly short-lived compared to the full dynamics, and because the Ohmic spectral density can be converged with the numerically exact reference method HEOM at a still-affordable $K=8$ and $L=2$ with $\delta t=0.01$~$\Delta^{-1}$. This is important because the 2-time results would otherwise be too computationally expensive to converge. 

The spin-boson model Hamiltonian, 
\begin{equation}\label{eq:Hsplit}
    \hat{H} = \hat{H}_{\rm S} + \hat{H}_{\rm B} + \hat{H}_{\rm SB}
\end{equation}
consists of system, bath, and system-bath terms. The system Hamiltonian has the form, 
%\vspace{-10pt}
\begin{equation*}
    \hat{H}_{\rm S} = \begin{bmatrix}
\varepsilon & \Delta \\
\Delta  & -\varepsilon
\end{bmatrix}.
\end{equation*}
The bath is composed independent harmonic oscillators,
\begin{equation}\label{eq:harm_bath}
    \hat{H}_{\rm B} = \frac{1}{2}\sum_{n} [\hat{p}_n^2 + \omega_n \hat{q}_n^2],
\end{equation}
which are antisymmetrically coupled to the two sites,
\begin{equation}
    \hat{H}_{\rm SB} = \hat{\sigma}_z \sum_{n} c_n \hat{q}_n \equiv \hat{\sigma}_z \otimes V_B, \tag{\ref{eq:Hsb}}
\end{equation}
where $\sigma_z$ is the $z$ Pauli matrix, $\hat{q}_n$ and $\hat{p}_n$ are the mass-weighted position and momentum operators for the $n^\rmm{th}$ harmonic oscillator in the bath, $\omega_n$ is the frequency of the $n^\rmm{th}$ oscillator and $c_n$ its coupling constant to the spin. In the spectroscopic initial condition we use in the Argyres-Kelley projector corresponds to an equilibrium bath described by its canonical density, $\rho_B = e^{-\beta \hat{H}_B}/ {\rm tr_B}{e^{-\beta \hat{H}_B}}$. 

The spectral density, $J(\omega) = \frac{\pi}{2}\sum_{n} c_n^2\delta(\omega - \omega_n)/\omega_{n}$, fully determines the system-bath coupling. We adopt the common Ohmic form with an exponential cutoff,
\begin{equation}
    J(\omega) = \eta \omega \e{-\omega/\omega_c},
\end{equation}
where $\eta$ is the usual Kondo parameter, $\lambda \equiv \eta \omega_c/\pi = \frac{1}{\pi}\int_0^{\infty} \dd{\omega}\ \frac{J(\omega)}{\omega}$ is the reorganization energy that quantifies the strength of system-bath coupling, and $\omega_c$ is the characteristic frequency of the bath.

To employ the mean-field Ehrenfest method to calculate correlation functions, we first perform a partial Wigner transform with respect to the bath variables, consistent with the derivation based on the quantum-classical Liouville equation \cite{Kapral2015a}, and then apply the standard time-dependent self-consistent field approximation under a classical treatment of the bath and a quantum treatment for the system \cite{McLachlan1964a, Stock1995a}. We adopt the protocol outlined in Ref.~\onlinecite{Montoya-Castillo2016}. Performing the partial Wigner transform of correlation functions for the rotated initial condition requires obtaining a Wigner transform of the commutator and anticommutator of the bath density, $\rho_B$, and the bath part of the system-bath coupling, $V_B$. For the linear term in Eq.~\ref{eq:K3bWignerEg}, acting the Poisson bracket on $\rho_{B, w}$ and $V_{B, w}$ returns $\rho_{B, w}$ and
\begin{equation}
    \zeta_{B,w} = -\sum_jc_jP_j\tanh{(\beta\omega_j/2)}/\omega_j,
\end{equation} 
which consists of a sum over bath momenta $P_j$, frequencies $\omega_j$, and coupling constants $c_j$.

For our Ehrenfest dynamics, we discretize the bath into $n_\rmm{osc}=300$ oscillators employing the following relations\cite{Makri1999c, Craig2005a, Montoya-Castillo2017a} 
\begin{subequations}
\begin{align}
    \omega_n &= -\omega_c \ln \Big[ \frac{n - \frac{1}{2}}{n_{osc}}\Big],\\
    c_n &= \omega_n \sqrt{\frac{2\lambda}{n_\rmm{osc}}}.
\end{align}
\end{subequations}

\section{Derivation of Semiclassical 2-time Kernel}\label{app:kernel}
\vspace{-10pt}
Here, we show how to derive a closure of the self-consistently expanded \textit{multitime} kernels that offers improvement in both efficiency and accuracy when using approximate semiclassical dynamics. For an explanation of the \textit{1-time} auxiliary kernels and Wigner transforms invoked in the main text, we refer the reader to Ref.~\onlinecite{Montoya-Castillo2016}. 
% Note that the counterpart to $\bm{\mc{K}}^{(3b)}(t)$,
% \vspace{-10pt}
% \begin{equation}
% \bm{\mc{K}}^{(3f)}(t) = (\bm{A}\vert \e{-\imi \mc{L}t} \mc{QL} \vert\bm{A}),
% \end{equation}
% is required below.

Before starting to manipulate the 2-time kernel, we show that $\mc{QL}|\mathbf{A}) \rightarrow \mc{L}_{SB}|\mathbf{A})$ and $(\mathbf{A}|\mc{LQ} \rightarrow (\mathbf{A}|\mc{L}_{SB}$. First, we perform the common splitting of the Hamiltonian into three parts (Eq.~\ref{eq:Hsplit}),
\begin{equation}
    \mc{QL}\vert \bm{A}) = \mc{QL}_S \vert \bm{A}) + \mc{QL}_B \vert \bm{A}) + \mc{QL}_{SB} \vert \bm{A}).
\end{equation}
Since $\mc{L}_S$ is a system operator
\begin{equation}
    \mc{QL}_S \vert \bm{A}) = \mc{Q}\vert\bm{A})\bm{Z} = 0,
\end{equation}
where $\bm{Z}$ is a static rotation between the system observables. Hence, the resulting generalized vector is part of the projector, $\mc{P}$, such that its product with $\mc{Q}$ annihilates the state as $\mc{PQ} = \mc{QP} = 0$. Since $\mc{L}_B$ is a bath operator it commutes with both the system operators, $\ket{j}\bra{k}$ and the canonical bath density, $\rho_B = e^{-\beta \hat{H}_B}/ \trb{e^{-\beta \hat{H}_B}}$. Hence, 
\begin{equation}
    \mc{L}_B \vert \bm{A}) = 0.
\end{equation} 
Finally, $\mc{Q}\mc{L}_{SB}\vert\bm{A})$ is expanded as $\mc{L}_{SB}\vert\bm{A})-\vert\bm{A})(\bm{A}\vert\mc{L}_{SB}\vert\bm{A})$ and we see immediately that the expectation of $\mc{L}_{SB}$ removes the second term since $\rho_B$ is quadratic and $V_B$ is linear, averaging to zero. The same logic up to the absence of any bath components holds for $(\bm{A}\vert\mc{LQ}$.

We can now obtain one of the key results in the paper: Eq.~\ref{eq:KsigmaWigner}. Starting from Eq.~\ref{eq:2timekernel}, we use the idempotency property, $\mc{Q}=\mc{Q}^2$, to bring down a $\mc{Q}$ to the right of the left-most $\mc{L}$. This enables us to maintain the correct recursive structure invoking the Dyson expansion\cite{Feynman1951} on the left propagator,
\begin{widetext}
    \vspace{-10pt}
    \begin{align}
        \bm{\mc{K}}^{\sigma_x}(t_2,t_1) &= (\bm{A}|\mc{LQ}\e{-\imi\mc{QL}t_2}\mc{Q}\hat{\sigma}_x\e{-\imi\mc{QL}t_1}\mc{QL}|\bm{A}) \\
        &= (\bm{A}|\mc{LQ} \e{-\imi\mc{L}t_2} \mc{Q}\hat{\sigma}_x \e{-\imi\mc{Q}\mc{L}t_1} \mc{Q}\mc{L} |\bm{A}) + \imi\int_0^{t_2}\ris\dd{s}\,(\bm{A}|\mc{LQ} \e{-\imi\mc{L}s} \vert\bm{A})(\bm{A}\vert \mc{L} \e{-\imi\mc{QL}(t_2-s)} \mc{Q}\hat{\sigma}_x \e{-\imi\mc{Q}\mc{L}t_1} \mc{Q}\mc{L} |\bm{A}) \label{eq:moreQ}\\
        &\equiv \bm{\mc{K}}^{\sigma_x}_L(t_2,t_1) + \imi\int_0^{t_2}\ris\dd{s}\, \bm{\mc{K}}^{(3b)}(s) \bm{\mc{K}}^{\sigma_x}(t_2-s,t_1).
    \end{align}
Thus, the 1-time auxiliary kernel, $\bm{\mc{K}}^{(3b)}$, appears where previously\cite{Sayer2024} one only required derivatives of the exact dynamics. We then invoke the Dyson expansion on the right-most propagator of $\bm{\mc{K}}^{\sigma_x}_L(t_2,t_1)$
    \begin{align}
        \bm{\mc{K}}_L^{\sigma_x}(t_2,t_1) &= (\bm{A}|\mc{LQ} \e{-\imi\mc{L}t_2} \mc{Q}\hat{\sigma}_x \mc{Q}\e{-\imi\mc{Q}\mc{L}t_1} \mc{Q}\mc{L} |\bm{A}) \\
        &= (\bm{A}|\mc{LQ} \e{-\imi\mc{L}t_2} \mc{Q}\hat{\sigma}_x \mc{Q}\e{-\imi\mc{L}t_1} \mc{Q}\mc{L} |\bm{A}) + \imi \int_0^{t_1}\ris\dd{s}\,(\bm{A}|\mc{LQ} \e{-\imi\mc{L}t_2} \mc{Q}\hat{\sigma}_x\mc{Q} \e{-\imi\mc{LQ}(t_1-s)} \mc{L}\vert\bm{A})(\bm{A}\vert \e{-\imi\mc{L}s} \mc{Q}\mc{L} |\bm{A})\\
        &\equiv \bm{\mc{K}}^{\sigma_x}_{LR}(t_2,t_1) + \imi\int_0^{t_1}\ris\dd{s}\,\bm{\mc{K}}^{\sigma_x}_L(t_2,t_1-s)\bm{\mc{K}}^{(3f)}(s),
    \end{align}
which removes the last remaining projected propagator. We insert the definition of $\mc{Q} = \mathbb{1} - \mc{P}$ for the two central complementary projectors and replace the outer instances that touch $\mc{L}$ with $\mc{L}_{SB}$ as planned,
    \begin{align}
        \bm{\mc{K}}^{\sigma_x}_{LR}(t_2,t_1) &= (\bm{A}|\mc{LQ} \e{-\imi\mc{L}t_2} \hat{\sigma}_x \e{-\imi\mc{L}t_1} \mc{Q}\mc{L} |\bm{A}) - (\bm{A}|\mc{LQ}\e{-\imi\mc{L}t_2}|\bm{A})(\bm{A}|\hat{\sigma}_x\e{-\imi\mc{L}t_1}\mc{QL}|\bm{A}) \nonumber\\
        &~~~~~-(\bm{A}|\mc{LQ}\e{-\imi\mc{L}t_2}\hat{\sigma}_x|\bm{A})(\bm{A}|\e{-\imi\mc{L}t_1}\mc{QL}|\bm{A}) + (\bm{A}|\mc{LQ}\e{-\imi\mc{L}t_2}|\bm{A})(\bm{A}|\hat{\sigma}_x|\bm{A})(\bm{A}|\e{-\imi\mc{L}t_1}\mc{QL}|\bm{A}) \\
        &= \bm{\mc{K}}^{(1)}(t_2,t_1) - \bm{\mc{K}}^{(3f)}(t_2){}^{\sigma_x}\bm{\mc{K}}^{(3b)}(t_1) - \bm{\mc{K}}^{(3f)\sigma_x}(t_2)\bm{\mc{K}}^{(3b)}(t_1) + \bm{\mc{K}}^{(3f)}(t_2)\bm{\Sigma_x}\bm{\mc{K}}^{(3b)}(t_1), \label{eq:finalKLR}
    \end{align}
where 
\begin{equation}
     \bm{\mc{K}}^{(1)}(t_2,t_1) \equiv (\bm{A}|\mc{L}_{SB}\e{-\imi\mc{L}t_2}\hat{\sigma}_x\e{-\imi\mc{L}t_1}\mc{L}_{SB}|\bm{A}).
\end{equation}
Here, the superscript $\sigma_x$ denotes modified kernels where the measurement occurs at the start or end of the propagation, as defined in the equation. Note that all the $\bm{\mc{K}}^{(3f)}$ and $\bm{\mc{K}}^{(3b)}$ that appear here also have the $\mc{L}_{SB}$ substitutions. Only the first term of Eq.~\ref{eq:finalKLR} requires 2-time information. This object is the 2-time analog for the 1-time $\bm{\mc{K}}^{(1)}$,
\begin{align}
    \bm{\mc{K}}^{(1)}_{nm}(t_2,t_1) &= \frac{1}{2}\tr{ \left( [\sigma_z, \hat{A}_n^\dagger][V_B, \rho_B]_+ + [\sigma_z, \hat{A}_n^\dagger]_+[V_B, \rho_B] \right)\hat{\sigma}_x(t_1)\left(V_B [\hat{A}_m, \hat{\sigma}_z] \right)(t_1+t_2) } \\
    &\equiv  \frac{1}{2}\tr{ \left( X_{ni}\hat{A}_i^\dagger[V_B, \rho_B]_+ + \imi Y_{ni} \hat{A}_i^\dagger[V_B, \rho_B] \right)\hat{\sigma}_x(t_1)\left(V_B \hat{A}_jX_{jm}  \right)(t_1+t_2) } \\
    &= \frac{X_{ni}}{2}\tr{\hat{A}_i^\dagger[V_B, \rho_B]_+\hat{\sigma}_x(t_1)\left(V_B  \hat{A}_j \right)(t_1+t_2)}X_{jm} \nonumber\\
    &~~~~~+ \imi\frac{Y_{ni}}{2}\tr{\hat{A}_i^\dagger[V_B, \rho_B]\hat{\sigma}_x(t_1)\left(V_B \hat{A}_j  \right)(t_1+t_2)}X_{jm} .
\end{align}
\end{widetext}
where the matrices $\bm{X}$ and $\bm{Y}$ are simple rotations from the terms involving $\sigma_z$.\cite{Montoya-Castillo2016} We use the commutator and anticommutator (denoted with subscript $+$) identity because this neatly separates the $V_{B,w}$ and $\zeta_{B,w}$ terms upon Wigner transform as Eq.~\ref{eq:KsigmaWigner} of the main text.

\onecolumngrid
\vfill
\pagebreak
\twocolumngrid
\subsection*{References}
\vspace{-10pt}
\bibliography{Postdoc}

\end{document}